# Size dependence of biexciton binding energy in strained ZnTe/(Zn,Mg)Te nanowire quantum dots

P. Baranowski[1], M. Zieliński[2], M. Szymura[1], N. Zawadzka[2], R. Pasławska[2], M. Wójcik[1], S. Kret[1], S. Chusnutdinow[1,3], P. Wojnar[1,3]

*1 Institute of Physic, Polish Academy of Sciences, 02-668 Warsaw, Poland*

*2. Institute of Physics, Faculty of Physics, Astronomy and Informatics, Nicolaus Copernicus University, ul Grudziądzka 5, PL-87-100 Toruń, Poland*

*3 Faculty of Physics, University of Warsaw, ul. Pasteura 5, PL-02-093 Warsaw, Poland*

*4 International Research Centre MagTop, Institute of Physics, Polish Academy of Sciences, 02-668 Warsaw, Poland*

**Abstract**

Nanowire quantum dots, i.e., heterostructures consisting of an axial insertion of low bandgap semiconductor within large band gap semiconductor nanowire, attract interest due to their emerging applications in the field of quantum communication technology. Here, we report on the fabrication of ZnTe/(Zn,Mg)Te nanowire quantum dots by molecular beam epitaxy and on a detailed investigation of the optical emission from individual structures by means of a combined study involving cathodoluminescence and micro-photoluminescence. A distinct dependence of the biexciton binding energy, defined as the spectral distance between the exciton and biexciton emission lines on the length of ZnTe axial insertions, is observed. With increasing dot length, not only does the biexciton binding energy value decrease distinctly, but also its character changes from binding to antibinding. The explanation of this effect relies on the appearance of a piezoelectric field along the nanowire axis, leading to a pronounced separation of electrons and holes. The change from a bound to an unbound character of biexciton energy can be well reproduced by theoretical calculations, which indicate an important contribution of excited hole states to this effect in the case of relatively large quantum dots.

**Introduction**

Over the last two decades, semiconductor nanowires (NWs) have been extensively investigated for their unique photonic [1–3], electronic [4,5] and thermal [6,7] properties. It is found that merging two or more semiconductors within a single NW may significantly extend their potential application range [8–10]. A prominent example of such NW-based heterostructure is the NW quantum dot, which is created by introducing a small axial insertion made of a low bandgap semiconductor into a NW with a larger bandgap [11–14] . As an effect, zero-dimensional optically active traps for electrons and holes are obtained [15,16], which attract the interest as an efficient single-photon source for applications in the field of quantum information technologies [17]. The advantage of the NW geometry relies on the increase of the extraction efficiency of the optical emission from the dots due to the waveguide effect and the suppression of the total internal reflection [16,18]. Additionally, placing a quantum dot inside NW allows for exact positioning of the quantum emitter and opens the possibility of stacking several quantum dots one upon each other [15], which may be used for the scopes of optical quantum computation [19].

On the other hand, NWs composed of crystals with non-central symmetry exhibit a piezoelectric effect when the stress is applied in the appropriate direction [20]. In particular, the c-plane in wurtzite- and the (111) plane in zinc-blende semiconductors are well-known to exhibit a non-vanishing polarity. At the same time, they are usually the front planes of the NWs [21]. Therefore, the application of the stress in the axial direction of NW usually results in the appearance of a piezoelectric field. The latter effect has opened a new research area, known as piezo-phototronics or piezotronics, which takes advantage of an effective manipulation of carrier transport and photoelectric properties by the internal electric field [22].

The piezoelectric effect may also impact the optical emission from individual quantum dots, as has been demonstrated for epitaxial quantum dots [23–25]. The presence of an internal electric field has been demonstrated to result in electron-hole spatial separation within planar wurtzite GaN/AlN quantum dots [23], elongated InAs/InP quantum sticks [24], and InAs/GaAs quantum dots grown on unusually oriented (211) GaAs substrate [25]. The piezoelectric field usually has a direct impact on the biexciton (XX) emission, i.e., a multiexcitonic complex consisting of two electron-hole pairs. The piezoelectric field induces the electron-hole spatial separation given mostly by the dot size, leading to the formation of two electric dipoles with the moments oriented in the same direction. The resulting repulsive dipole-dipole interaction reduces the XX-binding energy, which is observed directly in optical emission spectra from individual dots by means of micro-photoluminescence (micro-PL) [23,26].

In this work, it is demonstrated that the piezoelectric effect also affects significantly the optical emission of individual NW quantum dots. The studied structures consist of ZnTe axial insertions within (Zn,Mg)Te NWs oriented in (111)B polar direction fabricated by gold-assisted molecular beam epitaxy. It is found that XX-binding energy depends strongly on ZnTe insertion length and changes its character from a bound to an unbound when increasing ZnTe length. Theoretical calculations involving the piezoelectric effect reproduce at least qualitatively the observed tendency and reveal that the proper description XX-biding energy in ZnTe/(Zn,Mg)Te NW quantum dots must include the influence of the hole excited states in the case of relatively large dots with closely spaced ground and excited states.

**Molecular beam epitaxy**

Nanowire heterostructures are grown in a system for molecular beam epitaxy equipped with Zn, Mg and Te atomic flux sources by employing a gold-catalyst-assisted vapor-liquid-solid (VLS) growth mechanism. A similar growth procedure for ZnTe/(Zn,Mg)Te core/shell NWs has been described elsewhere [27], whereas the details of the present growth process of ZnTe/(Zn,Mg)Te NW quantum dots are included in the Experimetal Methods section. The NW growth is performed on (111)-oriented silicon substrate with gold nano-droplets at the surface. The essential part of our structures consists of VLS grown (ZnMg)Te / ZnTe / (Zn,Mg)Te consecutive NW segments stacked axially on top of each other. The average length of the lowest (ZnMg)Te segment amounts to 600 nm and that of the top segment – to 150 nm. The length of ZnTe axial insertions is estimated to be 7, 14 and 30 nm in the case of sample A, B and C, respectively. Subsequently, the NW-cores are coated with (Zn,Mg)Te radial shells at relatively low substrate temperature at which the VLS growth is slower than the radial epitaxial growth. The reason for the presence of (Zn,Mg)Te shell is the passivation of surface states which results in the increase of optical emission intensity from NW quantum dots [27]. At the same time ZnTe/(Zn,Mg)Te lattice mismatch results in the appearance of an additional strain within ZnTe-insertions [28].

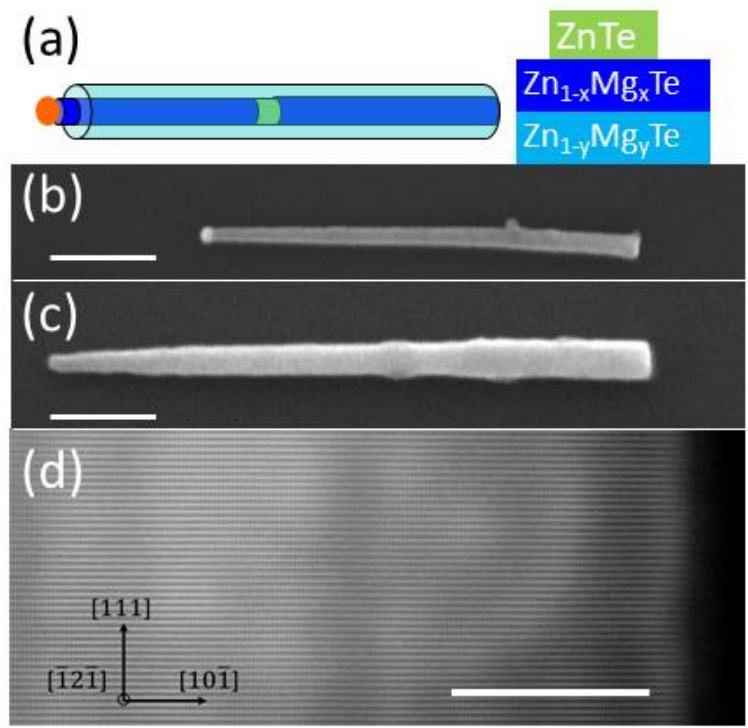

**Figure 1** Scheme of investigated ZnTe/(Zn,Mg)Te core/shell nanowire heterostructure grown by vapour liquid solid (VLS) mechanism (a), scanning electron microscopy of a typical (Zn,Mg)Te/ZnTe/(Zn,Mg)Te nanowire core only (b), and a typical core/shell structure (c). Acceleration voltage is 5 kV and the probe current - 40 ps. The scale bare corresponds to 200 nm (d) High resolution transmission electron microscopy of an selected ZnTe/(Zn,Mg)Te nanowire quantum dot heterostructure evidencing almost perfect crystalline structure in the core and in the shell region, acceleration voltage is 300 keV, scale bare corresponds to 10 nm

A scheme of the investigated NW heterostructure is presented in Figure 1a. Mg concentration, determined by Mg/Zn flux ratio, is expected to be similar in (Zn,Mg)Te cores and in (Zn,Mg)Te shells despite of the different growth mechanisms, VLS vs. epitaxial, used for growth of the respective NW parts [29]. Moreover, ZnTe/(Zn,Mg)Te semiconductor system is characterized by type I band alignment. Therefore, electrons and holes are expected both to localize within ZnTe axial insertion. A typical NW-core is compared to a typical core/shell structure in Figure 1b and 1c, respectively. It is found that the NW-diameter increases significantly upon the shell growth. The NW-core diameter measured at the half height varies from 25 nm to 45 nm, whereas the corresponding diameter of the core/shell structure – is from 80 nm to 100 nm, as found by statistics made on 200 NWs. By comparing the average values of the diameters, it is found that the average (Zn,Mg)Te shell thickness amounts to 24 ±8 nm. The NW crystalline structure is studied by high resolution transmission electron microscopy, Figure 1d. It is found that the NWs are always oriented in the [111] crystallographic direction. There are many NWs exhibiting an almost perfect crystalline structure of the core and the shell, whereas an epitaxial relation between these two parts is preserved. Elemental mapping by means of energy dispersive X-ray spectroscopy (EDS) reveals that the average Mg concentration in (Zn,Mg)Te nanowires amounts to 0.15. It is, however, not possible to determine reliably the exact sizes and positions of ZnTe insertions based on this study. However, the presence of ZnTe NW quantum dots is unambiguously confirmed by photoluminescence (PL) and cathodoluminescence (CL), as described in the next section.

**Identification of the optical emission from NW quantum dots**

In Figure 2a, low-temperature PL spectra from samples A, B and C containing ZnTe/(Zn,Mg)Te NW quantum dots with three different average lengths and from the reference (Zn,Mg)Te NWs without any insertion are presented. In the NW quantum dot containing samples PL spectrum consists of two emission bands. The high energy band is attributed to the emission from NW cores, whereas the addition of ZnTe axial insertions results in the appearance of an emission band at lower energies. In the case of sample C with the longest ZnTe axial insertions, the high energy emission line appears at 2.375 eV and the low energy line at 2.307 eV. When decreasing the average length of ZnTe axial insertion, the low energy line shifts toward higher energies and amounts to 2.317 eV and 2.350 eV for samples B and A, respectively. This energy shift is interpreted in terms of the quantum size effect, i.e., increasing the carrier confinement with decreasing the length of axial ZnTe insertion. On the other hand, the NW-core-related emission has always the maximum in the 2.375 eV – 2.385 eV range. We do not expect significant changes of this emission energy since the NW cores are always grown in the same conditions. The small variation of the emission energy of the high energy line in different samples might be attributed to the small changes in Mg flux during the growth of (Zn,Mg)Te NW cores.

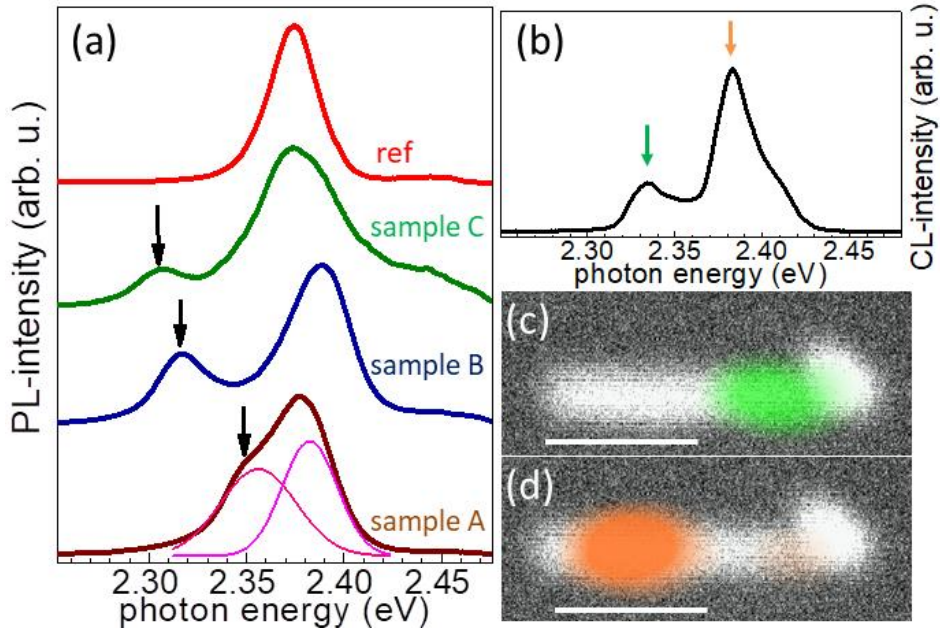

**Figure 2** Identification of the optical emission from ZnTe/(Zn,Mg)Te NW quantum dots (a) Photoluminescence from samples with different average lengths of ZnTe axial insertions. Arrows indicate emission lines from ZnTe NW quantum dots. For sample A, two Gaussian functions are fitted (cyan) to separate the emission from NW quantum dots and NW-cores. Excitation is performed with a 405 nm laser line, and the temperature of the measurement is 7 K. (b) Cathodoluminescence (CL) spectrum from a single ZnTe/(Zn,Mg)Te nanowire coming from sample A (c) CL-map performed at 2.340 eV and (d) at 2.390 eV superimposed with SEM image of the investigated NW. The probe current is 500 pA, acceleration voltage 15 kV and measurement temperature 7 K.

CL study allows us to identify the spatial position of the emitting object and assign unambiguously the emission lines to the NW-core or NW quantum dot emission. CL spectrum from an individual NW-heterostructure coming from sample A, i.e., with the shortest ZnTe axial insertion (7 nm on average), is presented in Figure 2b. It consists of two emission lines at 2.34 eV and 2.39 eV. In order to determine from which part of the NW heterostructure these emission lines are originating CL-maps are performed at 2.340 eV and 2.390 eV are presented in Figure 2c and 2d, respectively. It is found that in the case of the emission at 2.390 eV, the most intensively emitting area (orange) corresponds to the bottom NW part, whereas the emission at 2.340 eV is placed close to the top of the NW (green). This observation is consistent with the interpretation that the high energy line originates from NW-core and the low energy line comes from ZnTe axial insertion, which is expected to be placed within the top half of the NW. In the case of the emission at 2.390 eV, we note that it is present not only at the bottom of the NW but also at several other areas along the NW axis, but its intensity is significantly smaller there. The latter effect can be explained by the inhomogeneous thickness of the passivation shell. The thickest passivation shell is expected to be present at the bottom of the NW, so we expect the most intensive optical emission from this NW part. Most important conclusion from this study is, however, that CL investigation lead us to ascribe definitely the low energy line to ZnTe axial insertions and the high energy line to (Zn,Mg)Te NW-cores.

**Biexciton binding energy**

Micro-PL is used to investigate further the optical emission from individual NW heterostructures. A typical spectrum consists of two emission lines similar to the spectra measured by CL, Figure 2b. Therefore, we attribute the high energy line to the excitonic emission from (Zn,Mg)Te NW-core and the low energy line to the excitonic emission from ZnTe insertion. An interesting effect is observed when increasing the excitation fluence. One observes, namely, the appearance of an additional emission line located spectrally in the vicinity of the low energy line. In Figure 3a, the results from relatively short ZnTe insertion coming from sample A and in Figure 3b - from relatively large ZnTe insertion from sample C are presented. The main difference between these two structures is that the additional line appears either at lower or at higher energy than the excitonic emission from ZnTe-insertion in the case of Figure 3a and 3b, respectively. A closer look at these results reveals that the intensity of these additional lines increases superlinearly with the excitation fluence by following the power function with the exponent of 1.7 for both lines. On the other hand, the ZnTe-related emission line present in the low excitation regime increases linearly with the increasing excitation fluence. Based on this behavior (shown in the insets of Figure 3) we ascribe this line to single exciton emission, X, and the line appearing at high excitation fluence to the biexciton emission, XX. Importantly, the presence of XX emission is characteristic for low-dimensional structures and has not been observed neither in bulk ZnTe nor in ZnTe/(Zn,Mg)Te quantum wells nor in ZnTe/(Zn,Mg)Te core/shell NWs. Therefore, it may serve as indication of the zero dimensional confinement of charge carriers within ZnTe insertions.

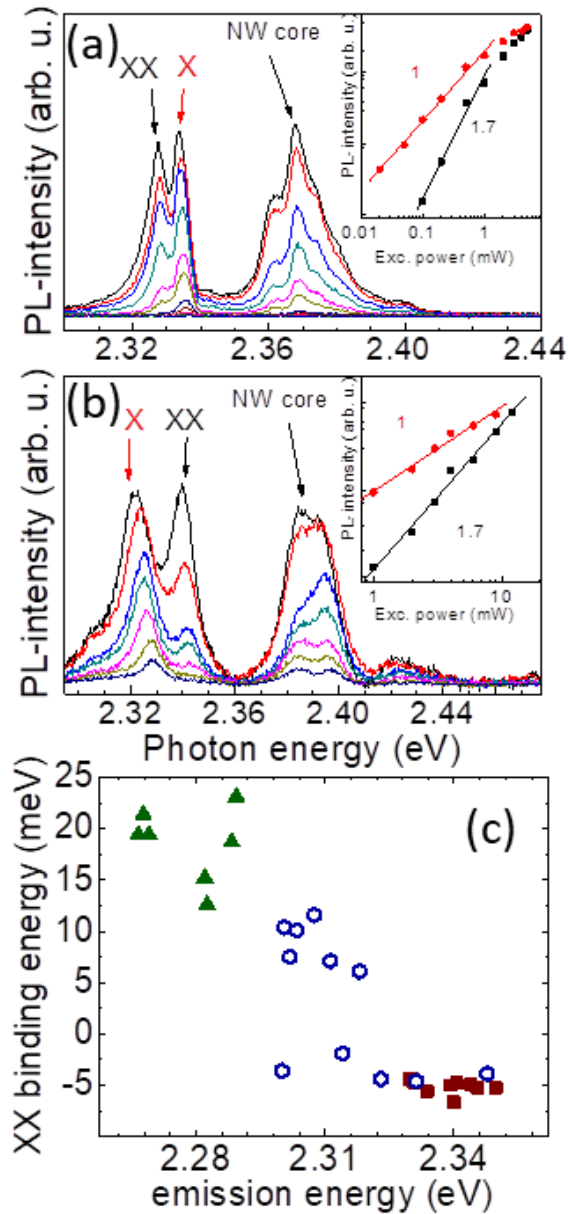

**Figure 3** Micro-photoluminescence spectra measured at different excitation fluencies from (a) sample A with relatively short (7 nm) ZnTe/(Zn,Mg)Te NW quantum dots and (b) sample C with relatively long (45 nm) NW quantum dots. Insets in (a) and (b) represent the emission intensity of X and XX as a function of excitation power. Biexcitons exhibit either binding or antibinding character depending on the length of ZnTe quantum dot. (c) XX-binding energy defined as XX - X spectral distance versus the X emission energy plotted for several individual NW quantum dots with different lengths revealing a change from bound to unbound character of biexcitons. Brown rectangles correspond to the NW dots from sample A, open circles – from sample B, green triangles - from sample C. The temperature of the measurement is always 7 K, and the excitation performed with 405 nm laser line.

The most intriguing part of our research concerns the biexciton binding energy, defined as the energy difference between XX and X lines. As already mentioned, it is found that this quantity can be either positive or negative depending on the length of ZnTe insertion. In order to investigate this tendency in a detailed manner, the XX binding energy is determined for 25 individual NW quantum dots and plotted as a function of the exciton emission energy. These results are presented in Figure 3c. One observes a clear tendency: XX emission from large quantum dots emitting at relatively low energy exhibits an antibinding character reaching absolute values even larger than 20 meV, whereas a binding character of XX emission is characteristic for small quantum dots emitting at relatively high energy. The increase of the XX binding energy on the X-emission energy is almost linear in the energy range from 2.26 eV to 2.35 eV, Figure 3c, changing the unbound to bound character of XX-emission at about 2.31 eV.

For our further considerations, it is important to link the X-emission energy to the length of ZnTe insertions. For this purpose, we note that the emission lines from sample C, i.e., with the longest average ZnTe insertions, are characterized by the emission energies at the lowest energies, i.e., in the range 2.26 eV - 2.29 eV. The lines from the medium-sized ZnTe insertions (sample B) appear in the 2.30 eV – 2.35 eV energy range and the shortest ZnTe insertions (sample A) - at the highest energy range, from 2.33 eV to 2.35 eV. This fact indicates that the X-emission energy is determined predominantly by the ZnTe insertion length. Of course, not only the length of ZnTe-insertions, but also other NW parameters, such as, the diameter of the NW, thickness of (Zn,Mg)Te shell, and Mg concentration in the core and the shell, impact also the X-emission energy from an individual NW quantum dot. The presence of these additional factors is the reason for the spread of XX-binding energies at a given X-emission energy observed in Figure 3c.

The straightforward explanation of the XX-binding energy variation with increasing ZnTe size relies on the presence of the piezoelectric field within the NW heterostructure. The NWs are oriented in the (111) direction, which implies the appearance of the piezoelectric field in the presence of strain caused by the lattice mismatch between ZnTe insertion and surrounding (Zn,Mg)Te [30]. The resulting piezoelectric field should be oriented parallel to the NW axis and induce electron-hole spatial separation limited by the length of ZnTe insertion. This effect may, in turn, directly impact the biexciton binding energy, similar to the case of epitaxial quantum dots [23–25]. Particularly promising in view of entangled photon generation would be the situation when the biexciton binding energy is close to zero [31,32]. Such emitters can be found in sample B when the emission energy is close to 2.31 eV.

**Theory**
In order to verify whether the piezoelectric effect may explain these experimental findings detailed theoretical calculations are performed. Since the exact knowledge of ZnTe quantum dot sizes and compositions, affected significantly by Mg-diffusion at the interfaces, is limited, the goal of this study is to determine at least qualitatively whether the change from bound to unbound

XX character with increasing ZnTe length could be induced by the piezoelectricity only and to find important parameters influencing the XX binding energy in these structures.

We have calculated a series of ZnTe quantum dots, with a diameter equal to 18 nm, embedded in $Zn_{0.85}Mg_{0.15}Te$ nanowires with a diameter equal to 30 nm. Atomistic modeling starts with finding atomic positions that minimize total elastic energy using Keating's valence force field method [33,34] and minimization of strain energy performed with the conjugate gradient method [35]. For ZnTe, we take parameters from [33]; for MgTe, we use parameters described in our earlier work [28]. For $Zn_{0.85}Mg_{0.15}Te$ alloy, we determine VFF parameters in the virtual crystal approximation by combing ZnTe and MgTe parameters (i.e. assuming 0.85 ratio of ZnTe and 0.15 of MgTe). Once atomic positions are found, the piezoelectric vector is calculated as each atomic site:

$$P_i = e_{ijk}\varepsilon_{jk} \tag{1}$$

where $e_{ijk}$ are piezoelectric coefficients and $\varepsilon_{jk}$ is a strain tensor. For zinc-blend lattice this formula reduces, by symmetry, to

$$\vec{P}(\vec{r}) = e_{14}(\vec{r})(\varepsilon_{yz} + \varepsilon_{zy}, \varepsilon_{xz} + \varepsilon_{zx}, \varepsilon_{xy} + \varepsilon_{yx})(\vec{r}) = e_{14}(\vec{r})\vec{\varepsilon}_{shear}(\vec{r}) \tag{2}$$

where $e_{14}$ is the piezoelectric modulus, typically obtained from experiments or ab-initio calculations. For ZnTe, $e_{14}$ is predicted to be relatively small and equal to 0.028 C/m². However, for alloyed (Cd,Mg)Te or (Cd,Zn)Te systems $e_{14}$ can increase by factor 5, even due to very small (approximately 1%) strains present in alloyed systems. Thus, due to large (5%) lattice mismatch between (bulk) ZnTe and MgTe, even small, i.e. 15%, contribution of MgTe (i.e. $Zn_{0.85}Mg_{0.15}Te$) would lead to small, yet non-negligible, 0.8% strain. Hence, in the following we will use $e_{14}$=0.14 C/m², that is five times that of ZnTe, which is a reasonable assumption in view of results presented in Ref. [36]. We note, however, that such piezoelectric coefficient is still smaller than, e.g. $e_{14} = 0.16$ for GaAs. Next, the piezoelectric charge is calculated as the divergence of piezoelectric vector:

$$\rho_P(\vec{r}) = -\nabla \cdot \vec{P}(\vec{r}) = -\left(\nabla e_{14}(\vec{r})\right) \cdot \vec{\varepsilon}_{shear}(\vec{r}) - e_{14}(\vec{r})\nabla \cdot \vec{\varepsilon}_{shear}(\vec{r}) \tag{3}$$

here we expanded the derivation, to emphasize that the piezoelectric charge can occur due to both the gradient of strain (the second term) and well as gradient of $e_{14}$ which also happens in our case, and as such cannot be neglected. Finally, the piezoelectric potential is obtained by solving the generalized Poisson equation:

$$\Delta V_P(\vec{r}) = \frac{\rho_P(\vec{r})}{\epsilon_0 \epsilon_S(\vec{r})} - \frac{1}{\epsilon_S(\vec{r})} \nabla V_P(\vec{r}) \cdot \nabla \epsilon_S(\vec{r}) \tag{4}$$

which we solve by successive-overrelaxation method. The values of ZnTe static dielectric constant reported in a rather broad range of values [37], here we take 10.3 after Ref [38]. The dielectric constants for MgTe reports are very limited, with values around approximately 7 [37], which, by interpolation, would equal to 9.8 for $Zn_{0.85}Mg_{0.15}Te$ alloy, thus very close to bulk ZnTe. In the following we assume its value to be 10.3 for both, the dot and the nanowire, effectively neglecting the second term in the generalized Poisson equation (4).

Next, we obtain single-particle electron and hole states with the empirical tight-binding $sp^3s^*$ approach accounting for spin-orbit interaction [39] and strain via the Harrison scaling law of hopping matrix elements [40]. For ZnTe tight-binding model, we used a set of empirical tight-binding parameters taken from [41]. For $Zn_{0.85}Mg_{0.15}Te$ alloy [28]; we developed a set of parameters starting to form pure ZnTe, and the shifting material band gap, via a shift of on-site (diagonal) energies of s orbitals, and additionally by accounting for a valence-band offset between core and shell materials [42,43]. Next, the excitonic spectra are calculated using the configuration-interaction method described in Ref. [44] with Coulomb and dipole matrix elements calculated using an approach presented in Refs. [44,45].

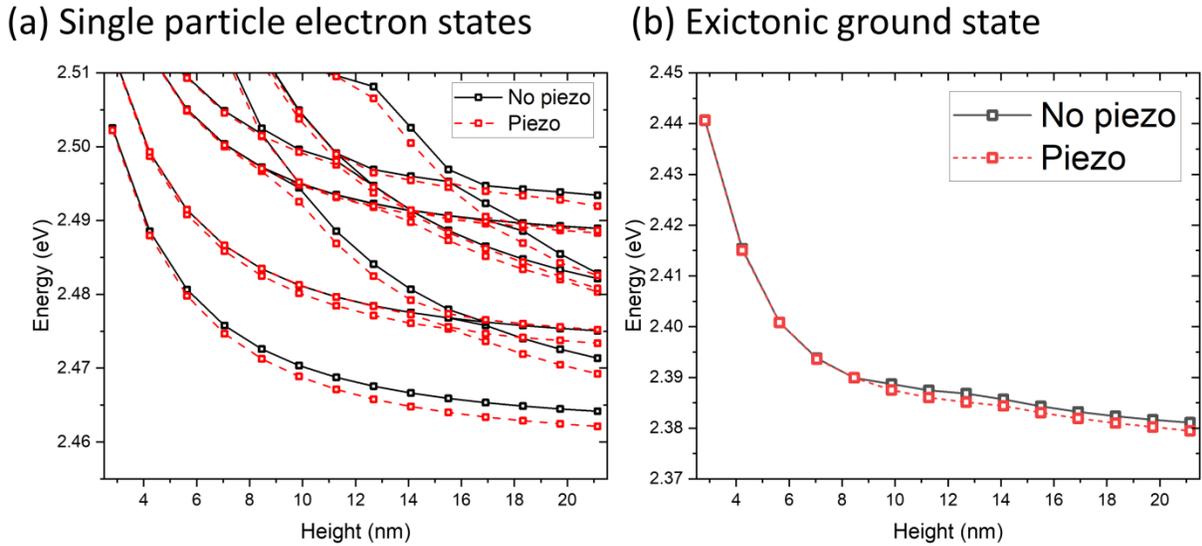

**Figure 4** Several lowest (a) single-particle electron state energies and (b) excitonic ground state energy calculated with piezoelectricity effect neglected or accounted for, and as a function of ZnTe section height.

Figure 4 shows several of the lowest single-particle electron state energies, with the piezoelectricity effect included or neglected as a function of ZnTe section height. The effect of piezoelectricity is rather weak, especially for shorter sections, and it basically leads to a moderate redshift of energies that increase with section height (as piezoelectric potential builds up with section length). Apart from the shift, the electron state reveals a shell-like structure (with s- and p-shells), a hallmark of two-dimensional (2D) confinement, for very flat sections until approximately 4 nm. With larger heights, a transition to a quasi-one-dimensional (1D) system occurs, with no apparent shell-structure. As the piezoelectric effect on the absolute single-particle energies is rather weak, so is the excitonic ground state energy, where the effect is notable only for ZnTe insertion heights larger than approximately 4 nm.

Piezoelectricity also has a rather small (on a scale of several meVs) effect of single-particle hole ground state energy, as shown in Figure 5. However, it has a rather pronounced effect on the spectra of excited hole states. In particular, for a case without piezoelectricity, there are apparent (h2-h1/h3-h2) level crossings at h=5.5 nm, which do not occur for the case with piezoelectricity. Moreover, when piezoelectricity is included, the hole ground state seems to be separated from the rest of the spectra, yet this energetic separation (in particular between h2-h1) decreases for larger heights.

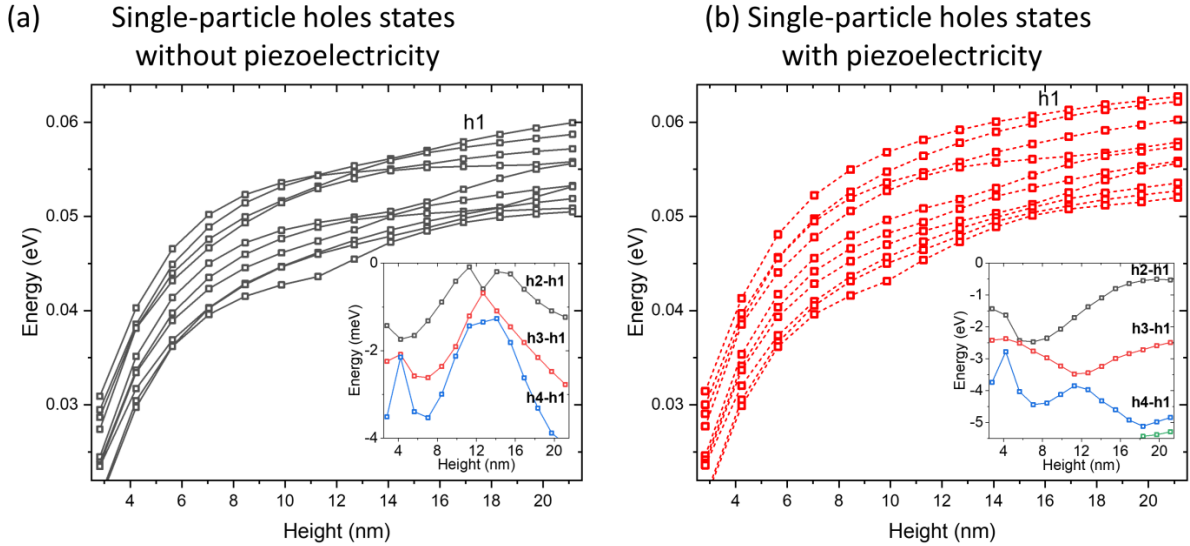

**Figure 5** Several lowest single-particle hole states with piezoelectricity effect neglected (a) or accounted for (b), and as a function of ZnTe section height. Insets show the details of hole spectra calculated as the energy difference between the three lowest excited (h2, h3, and h4) states and the ground state (h1).

To gain an insight into a rather complicated structure of hole states, Figure 6 shows the spatial distribution of the ground (ho1) and the first exciton (ho2) hole states charge densities (along the growth axis) for two ZnTe quantum dots of very different heights. There is a noticeable oscillation of charge densities on all plots due to the presence of atomic layers. Interestingly, for a small height not only the piezoelectricity affects ho1/ho2 charge densities, but also ho1 spatial distribution is very similar to ho2, which is due to strong confinement in the (z) growth direction. Here, ho1 is the s-like "node-less" wave function in all three spatial directions, whereas ho2 is of p-like character, yet in x/y-plane, (i.e., in the plane perpendicular to that on Figure 6), where it has a node, while there is no node in the z-direction. That makes ho1 and ho2 appear very similar to each other as viewed from the angle used in Figure 6. However, for a longer section, the confinement in the vertical direction is reduced, and for both cases (with and without piezoelectricity), the first excited state is of more complicated character, meaning it consists of two peaks with an apparent dip between them. (We emphasize, however, that since with study charge densities of a complex wave function, we are not able to determine the actual nodal structure of the ho2 wave function directly.) For a tall system, there are two apparent effects of piezoelectricity. One is the shift of the hole ground state as compared to the ho1 peak maxima for a flat system (Figure 6). A similar shift (in the opposite direction; not shown here) is also present for the ground electron state, and it leads to weak yet present electron-hole spatial separation due to the build-up of opposite sign piezoelectric charges at the bottom and top of the section [46]. The other effect is the pronounced change of ho2 spatial distribution, where one of its peaks becomes noticeably larger.

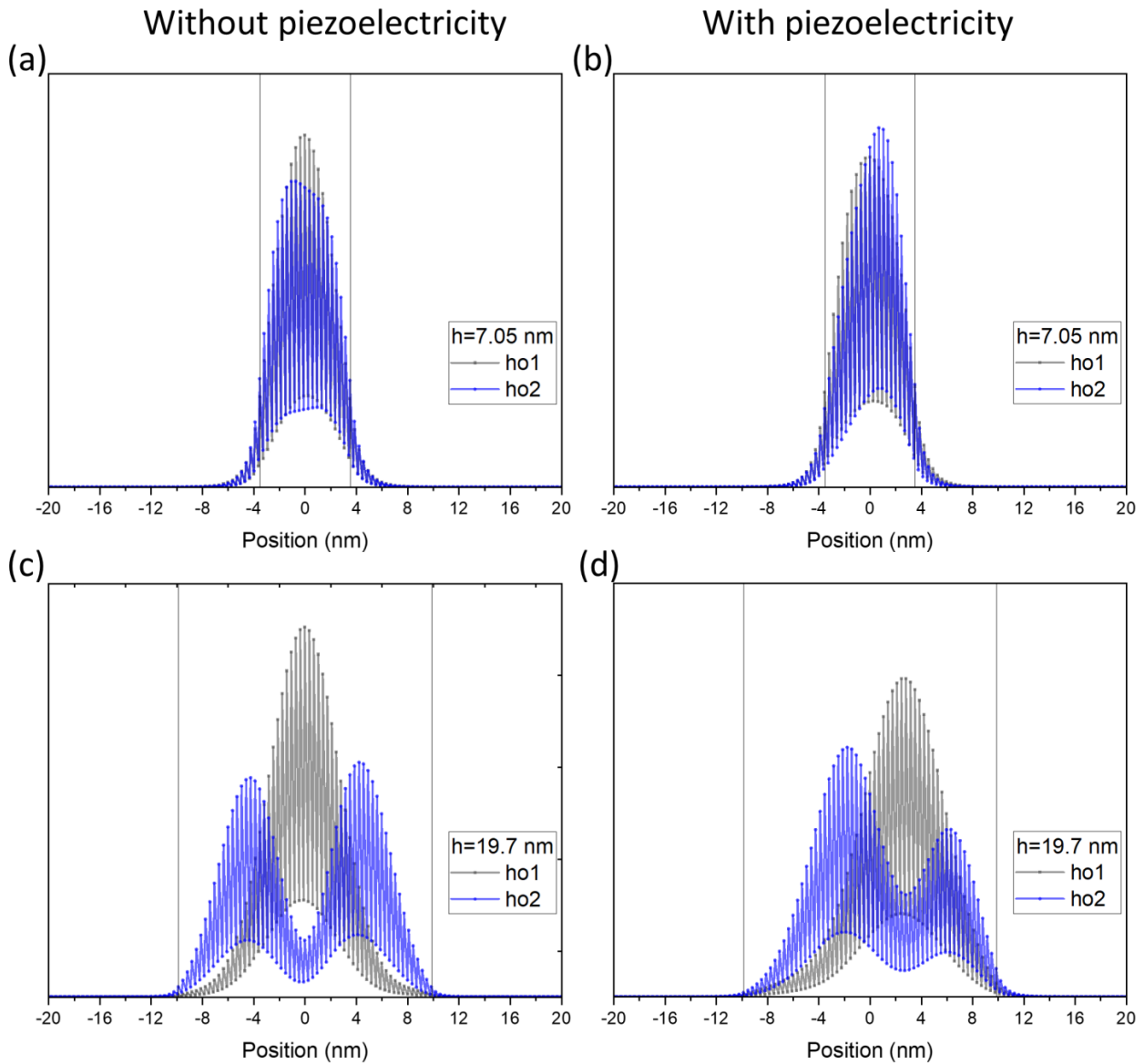

**Figure 6** Hole ground (ho1; gray) and first excited state (ho2; blue) charge densities with and without piezoelectric effected included in the calculation, for ZnTe quantum dots of different, i.e. height equal to 3.05 nm (a and b) and 8.54 (c and d). Note pronounced asymmetry of ho2 states for a tall (h=8.54 nm) section with piezoelectricity included, as well as shift of ho1 state maxima for the same case.

The increase of electron-hole spatial separation (between el1 and ho1 states) in the case with piezoelectricity included in the simulation could lead to the simultaneous decrease of electron-hole overlap. As such, it could manifest itself by, e.g., the decrease of electron-hole attraction and the decrease of exciton binding energy. However, no such large effect is visible in the excitonic spectra as studied earlier in Figure 4. Moreover, the incorporation of piezoelectricity actually decreased the excitonic energy (although mostly due to a red-shift of electron levels) or, in other words, increased the excitonic binding energy (as measured from the energy of a non-interacting electron-hole pair) thus the change in the electron-hole interaction due to piezoelectricity appears to be rather weak.

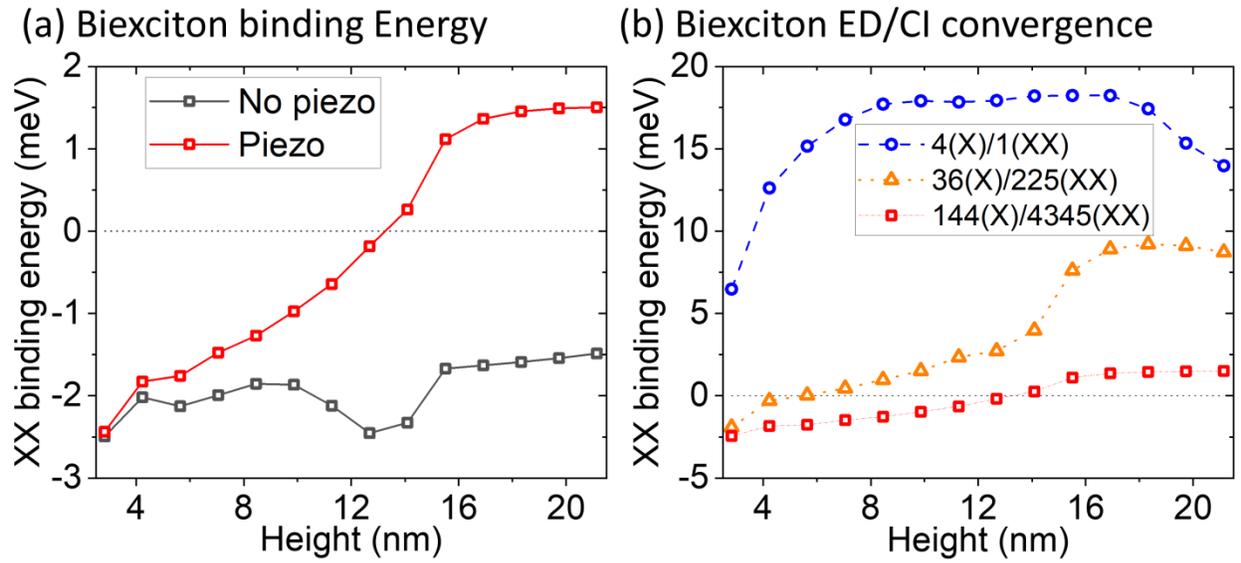

**Figure 7** Biexciton binding energy (a) with piezoelectricity effect neglected or accounted for (b), and as a function of ZnTe section height. Biexciton binding energy calculated with piezoelectricity included, yet for different level of approximation used in the CI calculation, where e.g. 144(X)/4345(XX) indicate 144 configurations used in the excitonic calculation and 4345 in the biexcitonic calculation.

The electron-hole spatial separation can, in principle, also strongly affect the XX binding energy. Therefore, Figure 7a shows the XX binding energy calculated with and without piezoelectricity included in the computation. Without piezoelectricity the XX binding energy only weakly depends on section height and it is negative, meaning the XX ground state energy is lower that the energy of two separated excitons. It is energetically beneficial for the system to confine two excitons, or in other words the XX is bound. With the inclusion of piezoelectricity nearly the same results are present, but only for the lowest considered heights. With the increase of height, however, the XX undergoes bound to unbound transition at about 5.5 nm where the single-particles hole-levels cross.

To further understand the origin of this process we study on Figure 7b the XX binding energy, this time with piezoelectricity, yet calculated at different levels of approximation used in the configuration interaction (CI) calculation. In particular, 4(X)/1(XX) is the calculation where only the ground electron and hole levels are included in the calculation, leading to 4 possible excitonic configurations, and only one XX configuration (with two electrons with different spin occupying the electron ground state and similar for holes). This level of approximation corresponds to applying an often-used formula for the XX binding energy: $\Delta XX = EXX - EX = Jee + Jhh - Jeh/2 - \Delta corr$, here $EXX$ and $EX$ are the XX and X ground energies correspondingly, $Jee$, $Jhh$, and $Jeh$ are electron-electron, hole-hole repulsion and electron-hole attraction integrals calculated in the respective ground states, where $\Delta corr$ is a correction due to correlations. For a single, 1(XX) biexciton configuration, and accounting for only 4 lowest X configurations, $\Delta corr$ is effectively neglected, and thus 4(X)/1(XX) strongly overestimates the overall XX binding energy. However, by inspecting this formula one expects that a separation between electron-hole would decrease the absolute value of $Jeh$, while simultaneously only weakly affecting $Jee$ and $Jhh$. Therefore, already at 4(X)/1(XX) level of approximation, the decrease of electron-hole separation explains (partially) increase the XX binding energy with the increasing section height. However, the piezoelectricity affects also excited states (in particular ho2), therefore Figure 7b shows also 36(X)/225(XX) case, which was obtained by including lowest 3 (6 with spin) electron and hole states leading to 36 possible electron-hole configurations for X and 225 configurations for XX. This calculation reveals the effect of piezoelectricity on ho2 states and, in turn, on the many-body properties. Finally, 144(X)/4345(XX) corresponds to lowest 6 (12 with spin) electron and hole states (corresponding to s-, p- and d-shell of typical, flat self-assembled quantum dot) leading to 144 and 4345 excitonic and biexcitonic configurations respectively. This number is of configurations is typically used as a tradeoff between CI convergence and computational trackability of the problem. Here, the inclusion of higher shells somewhat flattened the trend, but it did not change it. However, larger number of configurations and pronounced configuration mixing increased $\Delta corr$, and, as effect, decreased the XX binding energy, which is consistent with typical results obtained for quantum dots.

Summing up the theoretical considerations, it is found that the piezoelectric field within the investigated ZnTe/(Zn,Mg)Te NW heterostructure originates from the lattice mismatch between the dot and the barrier semiconductor, as well as from the mismatch of piezoelectric coefficients at the ZnTe/(Zn,Mg)Te interface which leads to the build-up of additional piezoelectric potential at the interfaces. The resulting piezoelectric field is oriented parallel to the NW axis and leads to the electron-hole spatial separation given by the size of the dot. This effect does not impact significantly the exciton ground state energy. It may have, however, a decisive impact on the XX binding energy. The straightforward picture is that the biexciton complex consists of two electron-hole pairs forming two electric dipoles with the moments oriented in the same direction. The longer ZnTe insertion the larger is the contribution of the repulsive dipole-dipole interaction leading to the overall decrease of XX binding energy. The calculations show that this picture must be complemented by the fact that piezoelectricity strongly affects also the excited holes states which is particularly important for relatively large quantum dots due to the relatively small spectral distance of ground and excited states. Implementing the excited states into the calculations in the frame of CI formalism leads to an important correction to XX binding energy values. However, it does not change the expected trend implying that the XX binding energy should increase with an increasing size of the dot.

**Conclusions**

In conclusions, optically active ZnTe/(Zn,Mg)Te NW quantum dots with excellent structural and optical properties are grown by molecular beam epitaxy employing VLS growth mechanism assisted with gold catalysts. The optical emission from ZnTe dots and (Zn,Mg)Te NW-cores are unambiguously identified by PL and CL measurements. It is found that the optical emission spectrum from an individual ZnTe insertion consists of X and XX lines which represents a strong indication for the zero-dimensional carrier confinement in this structure. Importantly, a distinct trend regarding the dependence of the XX-binding energy on ZnTe length is found. The XX-binding energy varies significantly from +7 meV to -23 meV and changes its character from bound to unbound when increasing ZnTe insertion length. The latter effect is explained by the presence of piezoelectric field which appears along the NW-axis and leads to electron-hole spatial separation limited by the dot length resulting in the reduction of the XX binding energy. Theoretical calculations demonstrate that the proper description of the XX binding energy values in the investigated system in the presence of piezoelectric field require taking into consideration the electron and hole excited states. The observed size dependence of the biexciton binding energy induced by the piezoelectric field within strained NW quantum dots heterostructures represents an important contribution to the general knowledge concerning the properties of NW quantum dots and can be applied for the design of efficient single photon sources based on NW heterostructures.

**Experimental Methods**

Molecular beam epitaxy system (EPI 620 system) is used for the growth of nanowire heterostructures by employing gold-catalyst assisted vapor-liquid solid growth mechanism. (111)-oriented silicon substrates with 1nm –thick gold layer are annealed at 700ºC which leads to the formation of gold/silicon eutectic alloy nano-droplets with the average diameter of 30 nm. Te, Zn and Mg atomic fluxes are characterized by the beam equivalent pressures of $4.5 \cdot 10^{-7}$, $2.0 \cdot 10^{-7}$ and $1.5 \cdot 10^{-8}$ mbar, respectively. The vapor-liquid-solid growth of (Zn,Mg)Te NW takes place at 380ºC for 35 min. The essential part of our structures consists of (ZnMg)Te / ZnTe / (Zn,Mg)Te consecutive NW segments. Three samples with different average lengths of ZnTe axial insertions are fabricated for the purposes of the present investigations. ZnTe segment is grown for 10, 20, and 60 seconds with the estimated average length of 7, 14 and 30 nm in the case of sample A, B and C, respectively. The top (Zn,Mg)Te segment is deposited during 5 min corresponding to the length of about 150 nm. Subsequently, the NWs are coated with (Zn,Mg)Te shells. For this purpose the same Zn, Mg and Te fluxes are used and the substrate temperature is lowered to 330 ºC. The shell growth takes place for 10 min and its estimated thickness amounts to 20 nm.

The crystalline structures of NW heterostructures is investigated using Titan Cube 80-300 transmission electron microscope (TEM) operating at 300 kV, equipped with an image corrector enabling high resolution imagining in the scanning transmission electron microscopy mode. The beam current is set to 60 pA, which corresponds to the size of the electron probe given by the full width at half maximum of the electron density distribution smaller than 0.13 nm.

For low temperature photoluminescence (PL), 'as grown' samples are placed in a closed cycle cryostat at 7 K. 405 nm laser excitation line is focused down to 0.1 mm diameter spot. To detect the signal 303 mm monochromator (SR 303i by Andor) equipped with a CCD camera is used.

Cathodoluminescence (CL) measurements are performed in Zeiss Ultra55 scanning electron microscope (SEM) equipped with a system for CL signal collection and detection (HCLUE by Horiba) and a helium-cooled sample stage (by Kammrath and Weiss). For these measurements, NWs are removed from the original substrate and dispersed on a clean Si substrate by using the sonication in an isopropanol bath. This procedure enables us to study the optical emission from an individual NW heterostructure.

For micro-photoluminescence (µ-PL) measurements NWs are excited by 405 nm laser beam focused onto a ~3 µm diameter spot by a microscope objective, which allows us to study the optical emission from individual quantum objects. The emitted light is collected by the same objective and detected by a 500 mm-monochromator (SR-500i by Andor) equipped with a CCD camera. The sample is placed on a cold finger inside a cryostat at 7 K. In order to resolve spatially the emission from individual NWs, the NWs from 'as grown' samples are dispersed on a clean silicon substrate. The average spatial distance between the NWs is significantly larger than the excitation spot size.

**Acknowledgements**

This work has been partially supported by the National Centre of Science (Poland) through grant 2017/26/E/ST3/00253, and by the Foundation for Polish Science project "MagTop" no. FENG.02.01-IP.05-0028/23 co-financed by the European Union from the funds of Priority 2 of the European Funds for a Smart Economy Program 2021–2027 (FENG). M.Z. acknowledges support from the Polish National Science Centre based on Decision No. 2018/31/B/ST3/01415